\documentclass[twocolumn,aps,prc,showpacs,floatfix]{revtex4}
\usepackage{graphicx}
\usepackage{dcolumn}
\usepackage{bm}

\begin{document}

\title{Decomposition of the sensitivity of the symmetry energy observables}

\author{He-lei Liu$^{1,2}$}
\author{Gao-Chan Yong$^{2}$}\email{yonggaochan@impcas.ac.cn}
\author {De-Hua Wen$^{1}$}
\affiliation{%
$^1${School of Sciences, South China University of Technology,
Guangzhou 510641, P.R. China}\\ $^2${Institute of Modern Physics,
Chinese Academy of Sciences, Lanzhou 730000, China}
}%

\date{\today}

\begin{abstract}
To exactly answer which density region that some frequently used
symmetry-energy-sensitive observables probe, for the first time,
we make a study of decomposition of the sensitivity of some
symmetry-energy-sensitive observables. It is found that for the
Au+Au reaction at incident beam energies of 200 and 400
MeV/nucleon, frequently used symmetry-energy-sensitive observables
mainly probe the density-dependent symmetry energy around
1.25$\rho_{0}$ (for pionic observables) or 1.5$\rho_{0}$ (for
nucleonic observables). Effects of the symmetry energy in the
low-density region is in general small but observable. The fact
that the symmetry-energy-sensitive observables are not sensitive
to the symmetry energy in the maximal baryon-density region
increases the difficulty of studying nuclear symmetry energy at
super-density.

\end{abstract}


\maketitle

\section{Introduction}

In the last 20 years, great progress has been made in the study of
a new branch of nuclear physics, i.e, the isospin nuclear physics
\cite{esym,a2005,b2005,c2008}. Theoretical studies have shown
that, within the parabolic approximation, the energy per particle
in asymmetric nuclear matter can be approximately expressed as
$E(\rho,\delta)=E(\rho,\delta=0)+E_{sym}(\rho)\delta^2$, where
$\delta\equiv(\rho_n-\rho_p)/(\rho_n+\rho_p)$ is the isospin
asymmetry parameter and $E_{sym}(\rho)$ is the density-dependent
nuclear symmetry energy. The latter has been studied for decades
due to its great importance in both nuclear physics and
astrophysics
\cite{apply1,a2005,b2005,apply3,apply4,c2008,betty1,betty2,betty3}.
Although significant progress has been made, the symmetry energy
is still subject to uncertainties especially at high-density
region \cite{guo2013,guo2014}. Nowadays, many sensitive
observables have been identified as promising probes of the
symmetry energy, such as the $\pi^-/\pi^+$ ratio \cite{NPDF2,
pion1,pion2,pion3,pion4,pion5,pion6}, energetic photon as well as
$\eta$ \cite{yong14,xiao14,yongph}, the neutron to proton ratio
$n/p$ \cite{np1,np2,np3}, $t/^3He$ \cite{t1,t2}, the isospin
fractionation \cite{frac1,frac2,frac3,np2} and the neutron-proton
differential flow \cite{NPDF1,NPDF3}. However, one only knows
these observables are in general sensitive to the high-density or
low-density behaviors of the symmetry energy at certain beam
energy whereas none knows the decomposition of sensitivity of the
symmetry energy observables in the whole density region. Such
knowledge surely affects obtaining information of the density-%
dependent symmetry energy from comparisons of theoretical
simulation and experimental data. In this study, based on the
isospin-dependent transport model, we address the above question.

\section{Modeling nuclear potential in the IBUU transport model}

\begin{figure}[t]
\centering
\includegraphics [width=0.5\textwidth]{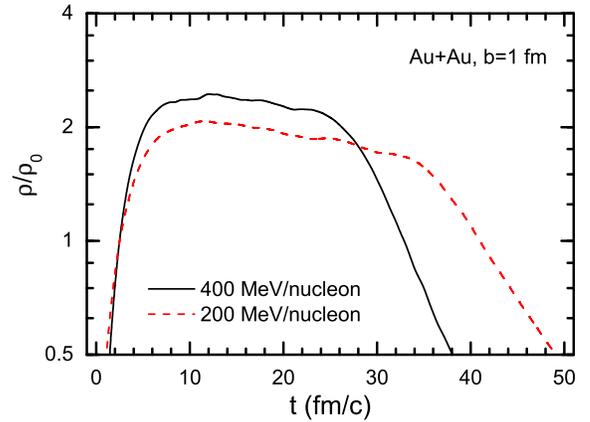}
\caption{\label{density} (Color online) Evolution of the central
baryon density in the reaction of $^{197}Au+^{197}Au$ at beam
energies of 400 and 200 MeV/nucleon with an impact parameter of
b=1 fm. $\rho_{0}$ denotes the nuclear saturation density.}
\end{figure}
\begin{figure}[t]
\centering
\includegraphics [width=0.5\textwidth]{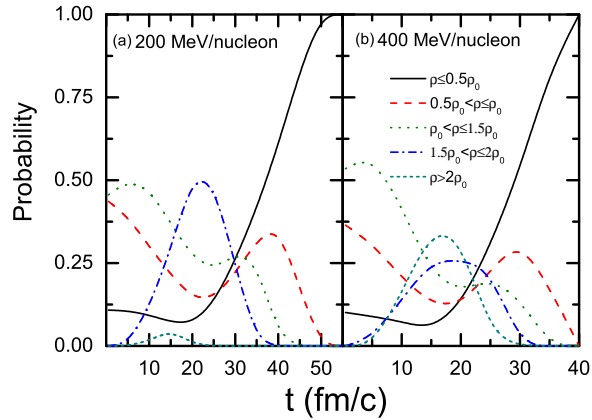}
\caption{\label{percentage} (Color online) Evolution of the
distribution percentage of baryon number in different density
regions in the central reaction of $^{197}Au+^{197}Au$ at beam
energies of 400 and 200 MeV/nucleon, respectively.}
\end{figure}
In this study, for simplicity, we use the isospin and
density-dependent single particle potential
\begin{equation}
U(\rho,\delta,\tau)=U_0(\rho)+U_{sym}(\rho,\delta,\tau),
\end{equation}
where the isoscalar potential reads
\begin{equation}
U_0(\rho)=-356 \rho/\rho_0+303 (\rho/\rho_0)^{7/6}.
\end{equation}
This soft nuclear isoscalar potential (SBKD) with $K_{0}$ = 200
MeV, was firstly introduced by Bertsch, Kruse and Das Gupta
\cite{u0}. For the isovector potential
$U_{sym}(\rho,\delta,\tau)$, we use the form \cite{usym}
\begin{eqnarray}\label{v0}
U_{sym}(\rho,\delta,\tau)=&&4\tau\delta(1.27+25.4u-9.31u^2+2.17u^3 \nonumber \\
&&-0.21u^4)-\delta^2(1.27+9.31u^2-4.33u^3\nonumber\\
&&+0.63u^4),
\end{eqnarray}
where $u=\rho/\rho_0$ is the reduced bayron density and $\tau$ =
1/2 (-1/2) for neutrons (protons). This symmetry potential roughly
stands for one of the frequently used symmetry potential in
nuclear transport models.

\section{Methods and Results}

To probe the density-dependent symmetry energy, it is instructive
to know the time evolution of central maximal baryon density
reached in heavy-ion collisions. Figure~\ref{density} shows the
evolution of the central maximal baryon density reached in
$^{197}Au+^{197}Au$ reaction at beam energies of 400 and 200
MeV/nucleon with an impact parameter of b=1 fm. One can see that
the central maximal baryon density reached is about 2.5$\rho_0$
for the incident beam energy of 400 MeV/nucleon and about
2.0$\rho_0$ for 200 MeV/nucleon case. And one can also see that
the supra-density nuclear matter exists longer for 200 MeV/nucleon
case than that for 400 MeV/nucleon case.

In order to better understand our target of studying the
sensitivity of symmetry-energy-sensitive observables in different
density regions, as shown in Figure~\ref{percentage}, we plot the
evolution of the distribution percentage of baryon number in
different density regions. We can see that the distribution
percentage of baryon number in different density regions changes
with reaction time. More baryons lie in the density region around
1$\sim$1.5$\rho_0$ in the whole reaction process. At the beginning
of the reaction (t $\geq$ 1 fm/c) nucleons in the two colliding
nuclei are slightly compressed, thus more nucleons lie in the
density region $\rho > \rho_{0}$. Immediately after this, the
central density of the reaction increases rapidly due to
compression as shown in Figure~\ref{density}.

Since baryons lie in different density regions, it is thus
necessary to study in which density region the
symmetry-energy-sensitive observables show maximal sensitivity. In
order to know in which density region the frequently used symmetry
energy sensitive observables show maximal sensitivity to the
symmetry energy, similar with the study in Ref.~\cite{liuy05}, in
the whole density region ($0 < \rho < 2.5 \rho_{0}$) we use the
$U_{sym}$ as the standard calculation, which gives a value of one
observable $R_{0}$, i.e.,
\begin{equation}
U_{sym}^{0 < \rho < 2.5 \rho_{0}} \rightarrow R_{0}.
\end{equation}
To see the relative sensitivity of this observable in different
density regions (i.e., $\rho_{1} \leq 0.5 \rho_{0},~
0.5 \rho_{0} < \rho_{2} \leq \rho_{0}, ~\rho_{0} < \rho_{3} \leq 1.5%
\rho_{0}, ~1.5\rho_{0} < \rho_{4} \leq 2 \rho_{0}, ~\rho_{5} > 2
\rho_{0}$ ), we turn off the symmetry energy in one density region
but keep the symmetry energy in the residual density region. We
thus get the other value of this observable $R_{i}$, i.e.,
\begin{equation}
U_{sym}^{0 < \rho < 2.5 \rho_{0}}- U_{sym}^{\rho_{i}} \rightarrow
R_{i} (i=1,2,3,4,5).
\end{equation}
Through comparing these new computational results with the
standard calculation $R_{0}$, one can obtain the relative
sensitivity in a certain density region, which reads
\begin{equation}
RS = \frac{|R_{0}-R_{i}|}{R_{0}} \times 100.
\end{equation}
In the following, we demonstrate decomposition of the sensitivity
of the frequently mentioned observables in the literature.

\begin{figure}[t]
\centering
\includegraphics [width=0.5\textwidth]{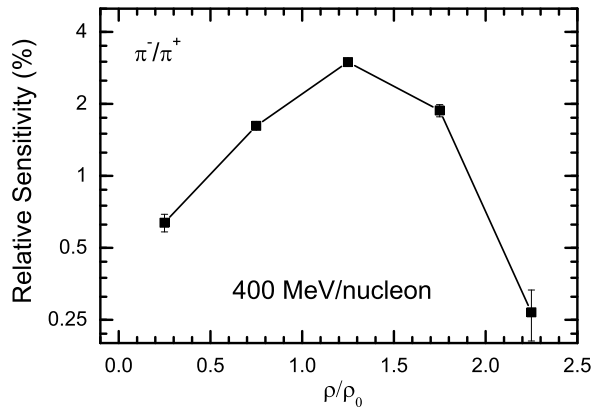}
\caption{\label{rpion400} Relative Sensitivity ( $RS$ ) of the
symmetry-energy-sensitive observable $\pi^-/\pi^+$ as a function
of density.}
\end{figure}
\begin{figure}[t]
\centering
\includegraphics [width=0.5\textwidth]{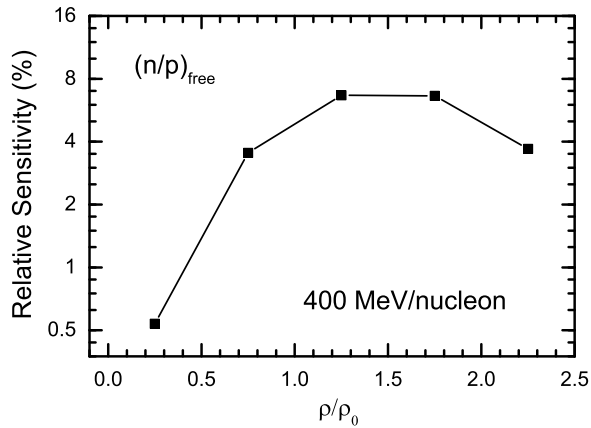}
\caption{\label{rnp400} Relative Sensitivity ( $RS$ ) of the
symmetry-energy-sensitive observable pre-equilibrium free n/p as a
function of density.}
\end{figure}
The $\pi^-/\pi^+$ ratio was first proposed as a probe of nuclear
symmetry energy in 2002 by \emph{Li} \cite{NPDF2}. It is generally
considered to be sensitive to the high-density behavior of the
symmetry energy. Shown in Figure~\ref{rpion400} is the
decomposition of the sensitivity of the symmetry energy observable
$\pi^-/\pi^+$ in different density regions. From
Figure~\ref{rpion400} we can clearly see that the maximal
sensitivity of $\pi^-/\pi^+$ to the density-dependent symmetry
energy lies in the density region around $1.25\rho_0$. Above
$1.5\rho_0$, sensitivity of $\pi^-/\pi^+$ to the symmetry energy
is roughly equal to that below $\rho_0$. With increase of density,
collision effect becomes larger. Thus one can see that the maximal
sensitivity does not lie in the maximal density region reached.
Due to rescatterings of pion meson in the low-density region,
charged pion ratio $\pi^-/\pi^+$ is also affected by the
low-density behavior of the symmetry energy in some degree. In
general, the sensitivity of charged pion ratio $\pi^-/\pi^+$ to
the nuclear symmetry energy is larger in the high density region $
\rho
> \rho_{0} $ than that in low density region $ \rho
< \rho_{0} $.

Shown in Figure~\ref{rnp400} is decomposition of the sensitivity
of the symmetry energy observable pre-equilibrium free neutron to
proton ratio n/p in different density regions. From
Figure~\ref{rnp400}, we can clearly see that the maximal
sensitivity of the free neutron to proton ratio n/p at the
incident beam energy of 400 MeV/nucleon lies in the density region
around $1.5\rho_0$. Because nucleon emission at pre-equilibrium of
the reaction does not suffer scatterings from the low-density
matter, the symmetry energy in the low density region thus has
minor effect.

\begin{figure}[t]
\centering
\includegraphics [width=0.5\textwidth]{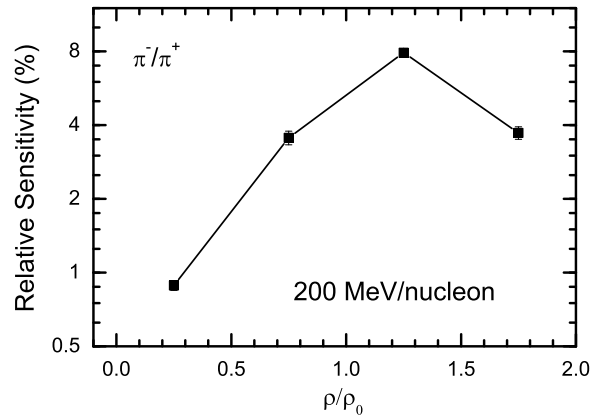}
\caption{\label{rpion200} Same as Figure~\ref{rpion400}, but at
the beam energy of 200 MeV/nucleon.}
\end{figure}
\begin{figure}[t]
\centering
\includegraphics [width=0.5\textwidth]{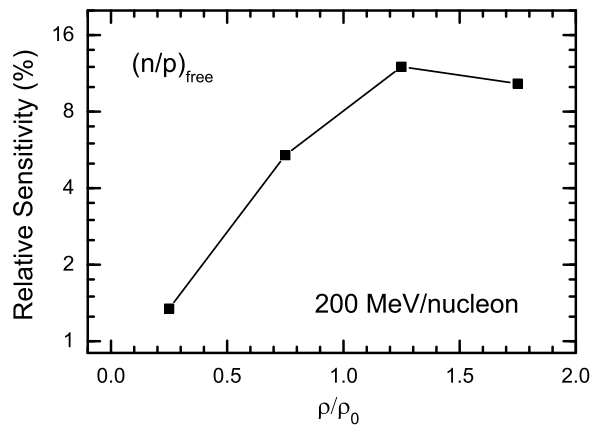}
\caption{\label{rnp200} Same as Figure~\ref{rnp400}, but at the
beam energy of 200 MeV/nucleon.}
\end{figure}
It is of interest to make similar study at incident beam energy
below 400 MeV/nucleon since such experiments of probing the
symmetry energy and related are being planned or performed at
RIKEN in Japan \cite{Lynch2014}. Shown in Figure~\ref{rpion200} is
decomposition of the sensitivity of the symmetry energy observable
$\pi^-/\pi^+$ in different density regions in Au+Au reactions at
the beam energy of 200 MeV/nucleon. We can see that the maximal
sensitivity of $\pi^-/\pi^+$ to the density-dependent symmetry
energy lies also in the density region around $1.25\rho_0$. And it
is in general still sensitive to the high density ($ \rho
> \rho_{0} $) behavior of
the symmetry energy. More interestingly, we can see that the
sensitivity of charged pion ratio $\pi^-/\pi^+$ to the symmetry
energy is about two times larger than that at 400 MeV/nucleon beam
energy. Therefore, one uses charged pion ratio $\pi^-/\pi^+$ to
probe the high-density behavior of the symmetry energy, it is
better to do experiments at relatively lower incident beam energy
\cite{zhang2012}. As for the observable pre-equilibrium free
neutron to proton ratio at incident beam energy of 200
MeV/nucleon, it is shown in Figure~\ref{rnp200}. The maximal
sensitivity of the pre-equilibrium free neutron to proton ratio
n/p at the incident beam energy of 200 MeV/nucleon also lies in
the density region around $1.5\rho_0$. In general, the symmetry
energy in the low density region has minor effect.

As shown in Figure~\ref{percentage}, because only a small
percentage of baryon lies in the maximum density region reached
and also in the maximum density there are larger collision
effects, symmetry-energy-sensitive observables are not sensitive
to the symmetry energy in the maximal baryon-density region. To
probe the symmetry energy at higher densities, one thus needs
heavy-ion collisions at even higher beam energy. However, the
larger collision effects at higher beam energy increase the
difficulty of studying nuclear symmetry energy.

\section{Conclusions}
In summary, within the isospin dependent IBUU transport model, we
studied the Au+Au reaction at incident beam energies of 200 and
400 MeV/nucleon. It is found that the symmetry-energy-sensitive
observables including charged pion ratio $\pi^-/\pi^+$,
pre-equilibrium free neutron-proton ratio $n/p$ mainly probe the
density-dependent symmetry energy around 1.25$\rho_{0}$ (for pion
emission) or 1.5$\rho_{0}$ (for pre-equilibrium nucleon emission).
Effects of the symmetry energy in the low-density region is in
general small but observable. Since the symmetry-energy-sensitive
observables are not sensitive to the symmetry energy in the
maximal baryon-density region, it is therefore a challenge to
probe the symmetry energy at higher densities.

\section{Acknowledgments}

This work is supported in part by the National Natural Science
Foundation of China under Grant Nos. 11375239, 11435014, 11275073
and the Fundamental Research Funds for the Central University of
China under Grant No. 2014ZG0036.

\end{document}